\documentclass[11pt]{revtex4}%
\usepackage[dvips]{graphics}%
\usepackage{subfigure}
\usepackage[dvips]{color}%
\usepackage{amsmath}
\usepackage{graphicx}%
\usepackage{amsfonts}%
\usepackage{amssymb}%
\usepackage{natbib}

\begin{document}
\title{Conformation characters of gel sheets with rotational symmetry: the role of boundary}
\author{Xiaobo Zhai\footnote{xbzhai8326@163.com}$^{a}$, Shengli Zhang$^{a,b}$ and Shumin Zhao$^{c}$}
\affiliation{$^{a}$Department of Applied Physics, Xi'an Jiaotong University, Xi'an 710049, China.\\
$^{b}$MOE Key Laboratory for Nonequilibrium Synthesis and Modulation of Condensed Matter, \\Xi'an Jiaotong University,Xi'an 710049, China.\\
$^{c}$Department of College Physics, Xi'an Jiaotong University, Xi'an 710049, China.}

\begin{abstract}
In this paper, we systemically study the conformation characters of
rotational symmetric gel sheets with free boundary and investigate the role
of boundary on the equilibrium conformation. In gel sheet the boundary
provides a residual strain which leads to re-distribution of stress and
impacts the shape of equilibrium conformation accordingly. For sheet with
boundary, the in-plane stretching energy is far larger than the bending
energy in some cases. It is intrinsic different from closed membrane. In
gel sheets, the boundary doesn't only quantitatively amend to the elastic
energy. The residual strain on boundary cooperates with bending and
stretching to determine the equilibrium conformation rather than just the
last two factors. Furthermore, on the boundary of gel sheet, there is an
additional energy induced by boundary line tension $\gamma $. If $\gamma =0$
, there is 10\% difference of elastic energy from the experimental
result. Finally, we discuss the effects of such line tension $\gamma $ and
propose a way to measure it by the border radius. It redounds to study the
physical origination of $\gamma $.
\end{abstract}

\maketitle

\section*{INTRODUCTION}

In thin biological tissues, the surface shapes are determined by their gene
 $[1-8]$ and external force $[9-11]$. The sheets with boundary have very
different shape from the closed membrane which has no boundary. For example,
the Gaussian curvature on boundary is negative, such as the leaves of
Acetabularia schenckii $[5]$; the vesicle with holes $[12]$; the leaves with
wavy edges which like torn edge of plastic garbage bags $[1]$. These
phenomena are general and can be found in other membrane such as the
synthetic gel sheets $[13]$. The rotational symmetric gel sheets have
brim-like boundary $[13,14]$, while the non-rotational symmetric gel sheets
are wavy-like $[13-16]$. These sheets also have negative Gaussian curvature
on boundary.

Theoretical studies reveal that the boundary of gel sheets have negative
Gaussian curvature $[17,18]$. The numerical simulations demonstrate that the
boundary regions of leaves with wavy shape and vesicle with positive Poisson
ratio all have negative Gaussian curvature near edges $[19,20]$, which
support our previous theoretical conclusion $[18]$. The similar results are
also found for the sheets with elliptic metric by E. Sharon {\em et al.},
who point out the boundary layers have abundant stretching energy $[21]$.
These discoveries imply that the sheet with free boundary has special
geometric shape and may have particular mechanical character, especially on
the range near its boundary.

On the boundary, because the molecules are difference between two sides of
boundary, there has a residual force. Then, boundary will have an additional
energy which can be described by a line energy density as the boundary line
tension $\gamma $ $[24,25]$. In many researches, people generally regard
that the changing of elastic energy arisen from $\gamma $ is weak. Therefore, in
theoretical studies $\gamma $ usually be omitted $[6,17,21-23]$.

In this paper, we investigate the role of boundary on reforming the
equilibrium shape and the distribution of strain for rotational symmetric
gel sheets, and study the effect of boundary line tension $\gamma $ on the
equilibrium conformation. Firstly, the boundary has a residual strain which
arises from the deformation to meet the boundary condition. This strain
leads to the geometric and mechanic characters in a sheet with boundary
obvious different from the sheet without boundary. Because of the residual
strain, the in-plane energy is enlarged near boundary. Then, the elastic
energy in a gel sheet with boundary is even two times larger than that in
the sheet without boundary. Consequently, the equilibrium conformation is
determined by the competition of bending, stretching and residual strain
rather than just the first two factors. Thus, in gel sheets, the boundary
doesn't just quantitatively amend to the elastic energy, but has essential
influence on the elastic energy.\ Secondly, $\gamma $\ will decrease the
border radius and elastic energy of gel sheet generally. If leaving out the
boundary line tension,\ the elastic energy has almost 10\% difference from
the experimental result. Thus, in studies of gel sheets, people need to
consider the boundary line tension. Finally, by the relationship between
 $\gamma $\ and equilibrium conformation, we propose a simple method to
measure boundary line tension by the border radius.

This paper is organized as follows: In Sec. II, the equilibrium shape
equations of gel sheets are deduced. It is found that the boundary has
special Gaussian curvature which is determined by Poisson ratio (Eqs. (6)).
In Sec. III, we study the physical effects of boundary. In closed membrane,
the stretching will reduce the conformation energy. However, in some cases,
the stretching increases the conformation energy because of the accumulation
of stretching energy near boundary. The stretching energy is even larger
than the bending energy. Then, we compare the sheet with and without
boundary and find the boundary has tremendous effect to the equilibrium
conformation. In Sec. IV, we research the relationship between the
equilibrium conformation and the boundary line tension. The boundary line
tension and the border radius have monotonic decreases relationship. Section
V is a conclusion.

\section{THE EQUILIBRIUM SHAPE EQUATIONS}
\label{sec:1}
In the gel sheets, we define two surface states, named as target state
 $\tilde{S}$ and equilibrium state $S$. The target state $\tilde{S}$\ is a
state which is in-plane strain free and it is an ideal conformation. The
equilibrium state $S$ is the final stable conformation of gel sheets.

The thin gel sheets can be represented as a 2D curved surface. For thin
sheets, the conformation energy can be written as $E=$ $E_{s}+E_{b}+E_{C}$
where $E_{s}$ is in-plane stretching energy, $E_{b}$ is bending energy and
 $E_{C}$ is an additional energy from the boundary line\ tension. The elastic
energy $F$ in the gel sheet is noted as $E_{s}+E_{b}$. The in-plane
stretching energy $E_{s}$ is represented by the displacement 2D vector field
${\bf u}$ on the surface of sheet which is determined by the difference
between equilibrium state $S$ and the target state $\tilde{S}$ $[18]$. The
bending energy $E_{b}$ is denoted as Helfrich $[26]$. The conformation
energy is $[26,27]$

\begin{equation}
\begin{split}
E &=\frac{1}{2}\int_{\tilde{S}}w_{s}d\tilde{A}+\int_{S}w_{b}dA+\oint_{C}\gamma
ds, \\
w_{s} &=2\mu u_{\alpha \beta }^{2}+\lambda u_{\alpha \alpha}^{2}, \\
w_{b} &=\frac{\kappa}{2}H^{2}+\kappa_{G}K,
\end{split}
\end{equation}
where $w_{s}$ and $w_{b}$ are the energy densities of in-plane stretching
energy and bending energy, $d\tilde{A}$ and $dA$\ are the area element on
the surface $\tilde{S}$ and $S$, respectively. $H$\ is the mean curvature
(we adopt it as the sum of principal curvatures rather than the average).
 $K$ is the Gaussian curvature.

For 2D surface, the elastic coefficients $\mu $, $\lambda $, $\kappa $\ and
 $\kappa _{G}$ satisfy $[28]$
\begin{equation}
\begin{split}
2\mu +\lambda &=\frac{hY_{0}}{1-\hat{\nu}^{2}}, \\
\lambda /(2\mu +\lambda ) &=\hat{\nu},  \\
\kappa /(2\mu +\lambda ) &=\frac{h^{2}}{12}, \\
\kappa _{G}/\kappa &=\hat{\nu}-1,
\end{split}
\end{equation}
where $h$ is the thickness of sheet, $Y_{0}$ is Young modulus and
$\hat{\nu}$ is the Poisson ratio.

For the sheet with a boundary $C$, the amount and type of molecules between
two sides of its boundary are different (gel's molecule is inside, and
environmental molecule is out side), so the boundary suffers a residuals
force which induce an additional energy on boundary. We can define a
boundary line tension $\gamma $, which represents the line density of
additional energy on boundary. This additional energy can be written as
 $E_{C}=\oint_{C}\gamma ds$\ $[24,25]$. In gel sheet, the shape of target
state is determined by the initial concentration, so the value of
 $\oint_{C}\gamma d\tilde{s}$ is constant. It can be regarded as a referenced
potential energy.\ Then, the boundary term of gel sheets is

\begin{equation}
E_{C}=\oint_{C}\gamma ds-\oint_{C}\gamma d\tilde{s}.
\end{equation}
where $ds$ and $d\tilde{s}$ are the line element on the boundary of
equilibrium state and target state, respectively.

For the surface $\tilde{S}$ and $S$ of rotational symmetric sheet, we choose
cylindrical coordinates $(\tilde{\rho},\varphi ,\tilde{z}(\tilde{\rho}
,\varphi ))$ and $(\rho ,\varphi ,z(\rho ,\varphi ))$ to describe them,
respectively. The in-plane strain tensor ${\bf u}$ is $[18,29]$

\begin{equation}
\begin{split}
u_{11} &=\frac{1}{2}[(\frac{\rho }{\tilde{\rho}})^{2}-1],  \\
u_{22} &=\frac{1}{2}[(\frac{dl}{d\tilde{l}})^{2}-1],   \\
u_{12} &=u_{21}=0.
\end{split}
\end{equation}
where $\tilde{l}=\int
\sqrt{1+\tilde{z}_{\tilde{\rho}}^{2}}d\tilde{\rho}$
and $l=\int \sqrt{1+z_{\rho }^{2}}d\rho $. If we denote $Y=-\frac{z_{\rho }}{%
\sqrt{1+z_{\rho }^{2}}}$, it has

\begin{equation}
\begin{split}
a &=\frac{Y}{\rho },  \\
c &=\frac{dY}{d\rho }.
\end{split}
\end{equation}
Substituting Eqs. (5) and Gauss-Bonnet formula $\int \int KdA=2\pi
-\oint_{C}k_{g}ds$ $[30]$ into Eqs. (1), and using the variational principle
$\delta F+\delta E_{C}=0$, we can obtain the equilibrium equations of
rotational symmetric sheets

\begin{equation}
\begin{split}
M_{1}-\nabla _{1}(M_{2}k_{g}^{-1})+\kappa \{\frac{1}{2}(H^{2}-4k_{n}^{2})+%
\frac{1}{2}\nabla _{1}[(H^{2}-4K)k_{g}^{-1}]\} &=0,  \\
aM_{2}+\kappa \lbrack \frac{1}{2}aH^{2}-\frac{F^{4}}{\rho
^{2}}\nabla _{1}(Hk_{g}^{-1})+H\frac{F^{4}}{\rho ^{2}}] &=0,
\end{split}
\end{equation}
and the sheets must satisfy the boundary conditions

\begin{equation}
\begin{split}
M_{2}+\frac{\kappa }{2}H(a-c)+\gamma k_{g} &=0, \\
c+\hat{\nu}a &=0.
\end{split}
\end{equation}
where $F=\sqrt{1+z_{\rho }^{2}}$, $\nabla _{1}=\frac{d}{dl}$, $M_{1}=\sigma
_{11}\frac{\rho }{\tilde{\rho}}\frac{d\tilde{l}}{dl}$, $M_{2}=\sigma _{22}
\frac{\tilde{\rho}}{\rho }\frac{dl}{d\tilde{l}}$, $k_{g}=\frac{1}{\rho F}$,
 $k_{n}=a$. $\sigma _{11}=(2\mu +\lambda )(u_{11}+\hat{\nu}u_{22})$, $\sigma
_{22}=(2\mu +\lambda )(u_{22}+\hat{\nu}u_{11})$. $\sigma _{11}$ and $\sigma
_{22}$\ are the stress along circumferential direction and radial direction,
respectively. The second equation of Eqs. (7) decides the relationship
between $\hat{\nu}$ and the sign of $c/a$, which also exists for
non-rotational symmetric sheets $[18,24]$.

We use reduction method to obtain the equilibrium equations (6). Comparing
with the previous work $[18]$, these equations are easily to be solved and
need lesser boundary condition. The second equation of Eqs. (7) is universal
in the studies of equilibrium shape. This equation will not be changed by
different variation method $[18,24]$. We note that the second equation of
Eqs. (7) determines the Gaussian curvature of boundary.

The conformations of gel sheets are determined by the equilibrium equations
(6) and boundary conditions (7) together. On the boundary, the sheets must
obey boundary conditions Eqs. (7) and need satisfy the equilibrium Eqs. (6).
It implies that the boundary also governs the equilibrium conformation
except for the bending and in-plane stretching. This is different from
closed membrane in which the equilibrium conformation is just determined by the
competition of bending and in-plane stretching. In this paper, we research the
open gel sheets with free boundary and discuss the effects of boundary by
comparing the sheets with and without boundary.

\section{CONFORMATION CHARACTERS INDUCED BY BOUNDARY}
\label{sec:2}
In biologic tissues, researchers observe that some thin rotational symmetric
membranes have negative Gaussian curvature on the boundary although the
Gaussian curvature is positive in the central regions, such as the leaves of
Acetabularia schenckii $[5]$ and the vesicle with holes $[12]$. The NIPA gel
sheets of dome-like shape also have negative Gaussian curvature on the
boundary $[13]$. The gel sheets which simulate growth of leaves have gradual
changed initial concentration $[13]$. Similarly as the growing process of a
leave, the shrinkage ratio in the artificial gel sheet can be regarded as
the growth. One can use shrinkage ratio $\eta _{0}$, which is linear with
initial concentration, to describe the target states as Ref. $13$ and Ref. $18$.

We classify the sheets by their target state's shape. The rotational
symmetric gel sheets have two types. In their target state, one has positive
Gaussian curvature and another has negative Gaussian curvature, such as
dome-like sheets and torus-like sheets. From the experiment $[13]$, we know
that the shape of target state is determined by $\eta _{0}$. The shrinkage
ratio $\eta _{0}$ of dome-like and torus-like sheets are $\sim r^{n}$
and $\sim r^{-n}$, respectively ($n>0$). To demonstrate the character
of Gaussian curvature on the boundary, we will investigate these two types
of gel sheets (if with no special annotation, the radius and thickness of
sheet are $r_{\max }=5$cm and $h=0.5$mm). Then, from these different sheets,
the boundary characters are summarized by the calculated data.

Furthermore, in order to show the conformation characters induced by the
boundary, for the gel sheet we compare its conformation in the equilibrium
state with boundary and without boundary, respectively. To deal with the
equilibrium state without boundary, we discuss an ideal infinite sheet and
calculate the equilibrium equations from inside radius and cut off at the
corresponding boundary $r_{max}$ (we doesn't attend the part beyond the
boundary at $r_{max}$).

In this section, we ignore the boundary line tension and will discuss it
detailedly in next section.

\subsection{Gaussian curvature}
From the boundary condition Eqs. (7), we derive that the surface must
satisfy $c/a=-1-\kappa _{G}/\kappa =-\hat{\nu}$ on the boundary. In
rotational symmetric sheet, it has $b=0$. The Gaussian curvature is
 $K=ac-b^{2}$, so the value of Gaussian curvature on boundary ($K|_{C}$) is
determined by the Poisson ratio $\hat{\nu}$. It can explain why on boundary
the sheets have negative Gaussian curvature in the experimental cases. If
 $\hat{\nu}>0$, on the boundary $a$ and $c$ have the opposite sign, so the
 $K|_{C}$ is negative. If $\hat{\nu}=0$, on the boundary Gaussian curvature is
zero. And if $\hat{\nu}<0$, the boundary has positive Gaussian curvature.
Generally the NIPA gel sheets have positive Poisson ratio $[31]$, so the
boundary of rotational symmetric gel sheets must has negative Gaussian
curvature.

Firstly, we study the dome-like sheets. To an experimental sheet (a
dome-like sheet in the Fig. 2b of Ref. $13$), we found that it really has
negative $K|_{C}$ by both the experimental data $[13]$ and our theoretical
calculation (Table 1). From the boundary condition Eqs. (7), we derive that
the $K|_{C}$ are determined by the sign of Poisson ratio as Fig. 1a. For
dome-like sheets with different target shape, in equilibrium state the
Gaussian curvature of them have similar distribution (Fig. 1). The increase
of initial concentration gradient just enlarges the extent of Gaussian
curvature (Fig. 1b). Furthermore, for different initial size and thickness
(we consider the thin thickness when the sheets can be bending), the
boundary also has the special geometric character: the absolute value of
 $K|_{C}$ is enhanced by the initial size increasing or the thickness
decreasing (Fig. 2). The distribution of Gaussian curvature is different in
equilibrium state and target state (Fig. 1a). For dome-like sheets, the
target states have positive Gaussian curvature. When $\hat{\nu}>0$, near
boundary the Gaussian curvature of equilibrium states is negative. When
 $\hat{\nu}<0$, the boundary of equilibrium state has positive Gaussian
curvature. For torus-like sheets, the Gaussian curvature of target state is
negative. When $\hat{\nu}>0$ the boundary of equilibrium state also has
negative Gaussian curvature. But when $\hat{\nu}<0$, the $K|_{C}$ is
positive. For both dome-like sheet and torus-like sheet, the $K|_{C}$
decrease when the Poisson ratio increases (Fig. 3). And Gaussian curvature
on the boundary depends on the Poisson ratio as $K|_{C}\sim \hat{\nu}^{2}$.

\begin{figure}
\centering
\renewcommand\figurename{\small Fig.}
\subfigure[]{
  \includegraphics[width=5cm]{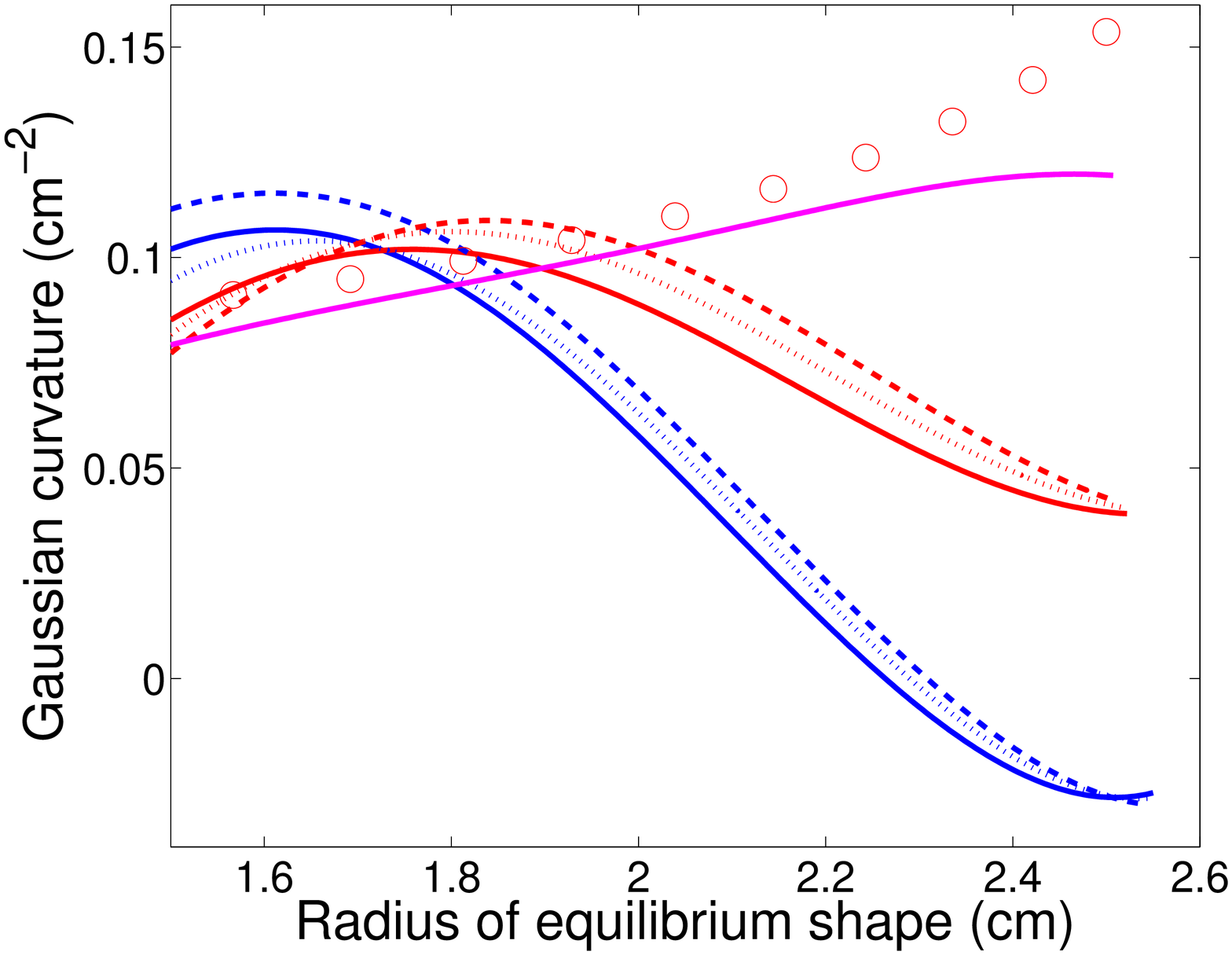}}
  \subfigure[]{
  \includegraphics[width=5cm]{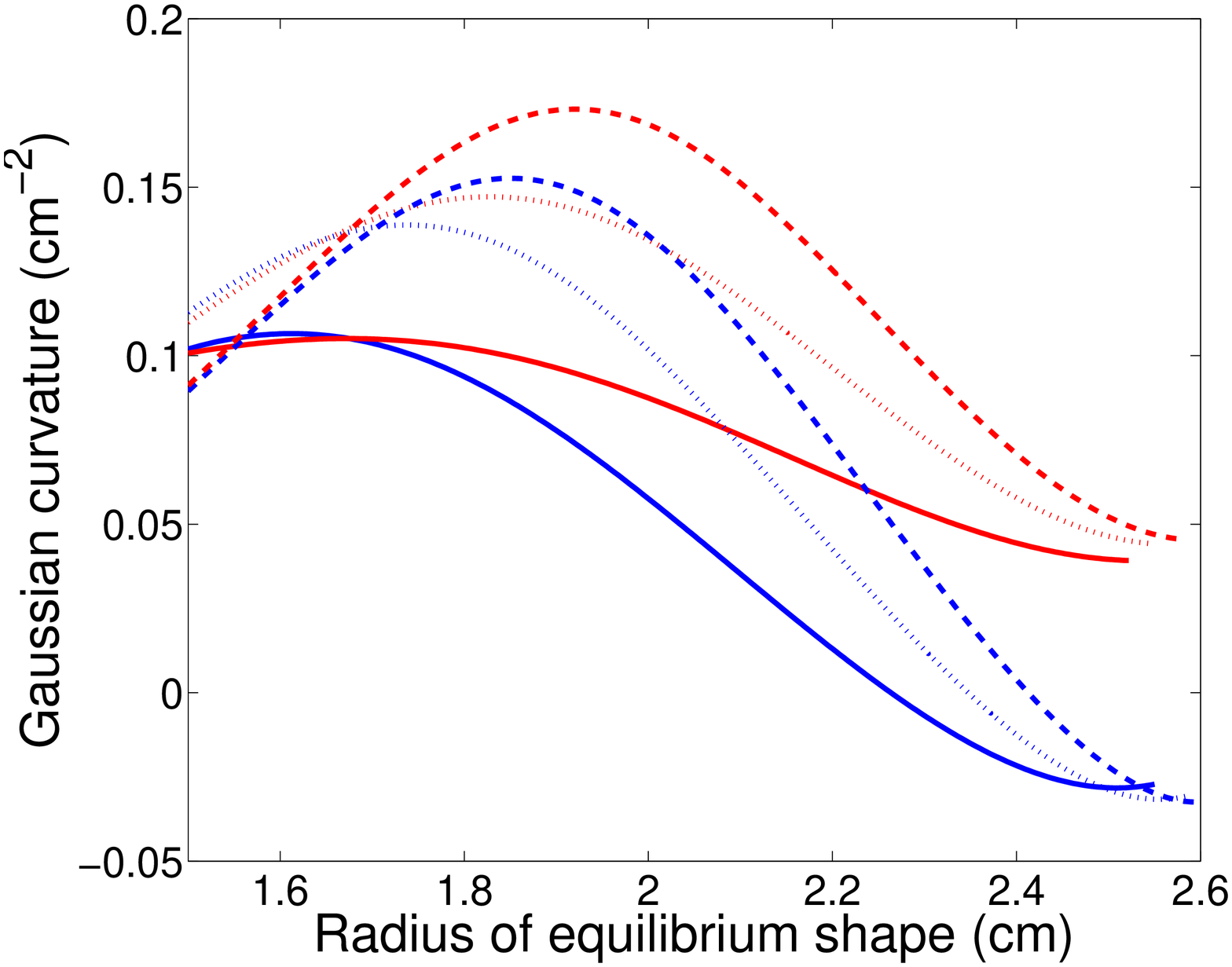}}
\caption{(a) The distribution of Gaussian curvature. The gel sheet has $\eta_{0}=0.6-r^{2}/250$. The hollow-dot line is in target state. The pink line is in the sheet without boundary when $\hat{\nu}=0.5$. The real line is in equilibrium state when $\hat{\nu}=0.5$\ (blue line) and $\hat{\nu}=-0.5$ (red line) to different boundary line tension, such as $\gamma /(2\mu+\lambda )=$\ $0$cm (real line), $0.0005$cm (dot-dash line), $0.001$cm (dashed). (b) The distribution of Gaussian curvature in different sheet. $\eta _{0}=0.6-r^{2}/250$ (real line), $\eta _{0}=0.6-r^{3}/1250$ (dot-dash line) and $\eta _{0}=0.6-r^{4}/6250$ (dashed) when $\hat{\nu}=0.5$ (blue line) and $\hat{\nu}=-0.5$ (red line).}
\end{figure}

\begin{figure}
\centering
\renewcommand\figurename{\small Fig.}
\subfigure[]{
  \includegraphics[width=4.5cm]{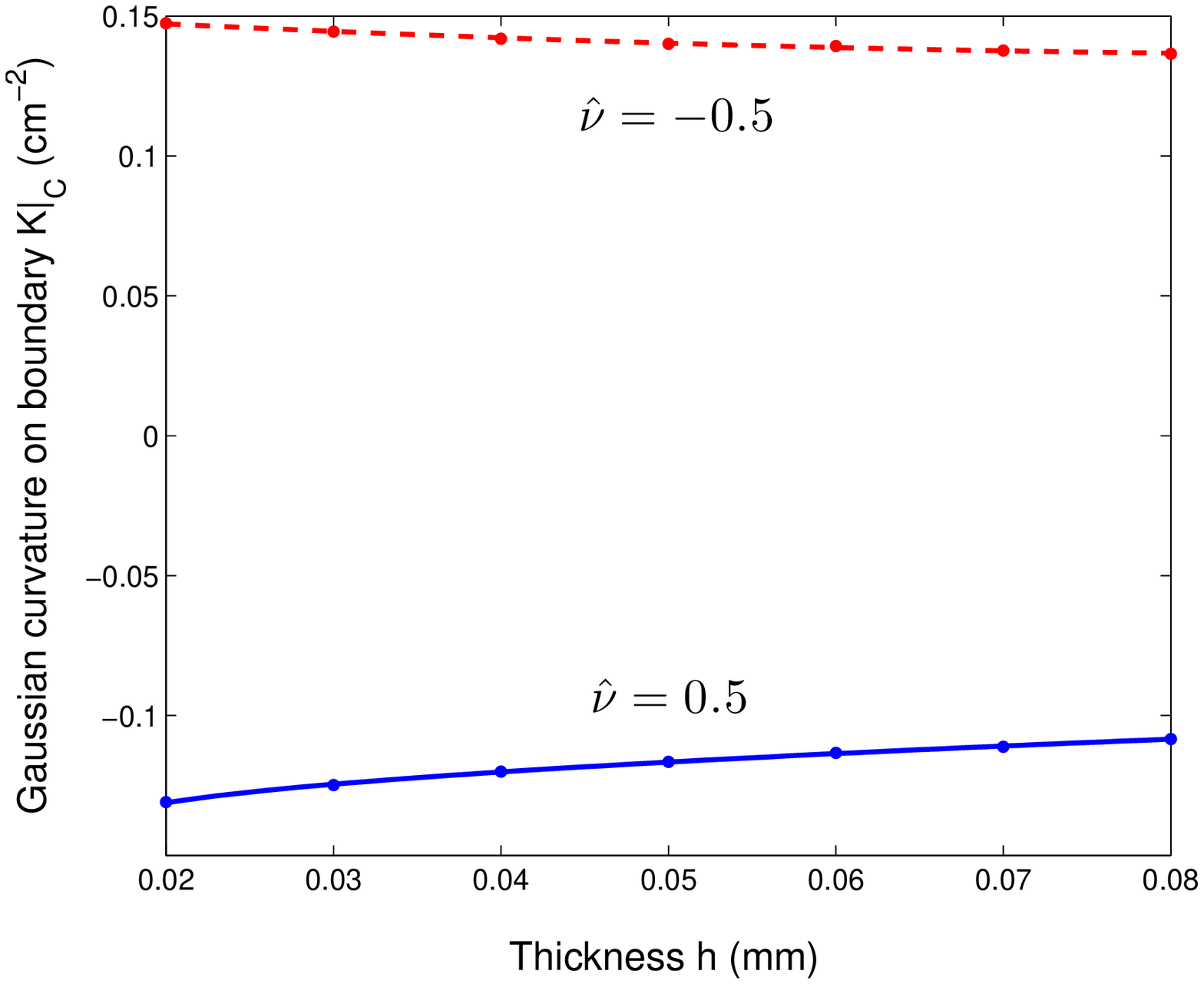}}
  \subfigure[]{
  \includegraphics[width=4.5cm]{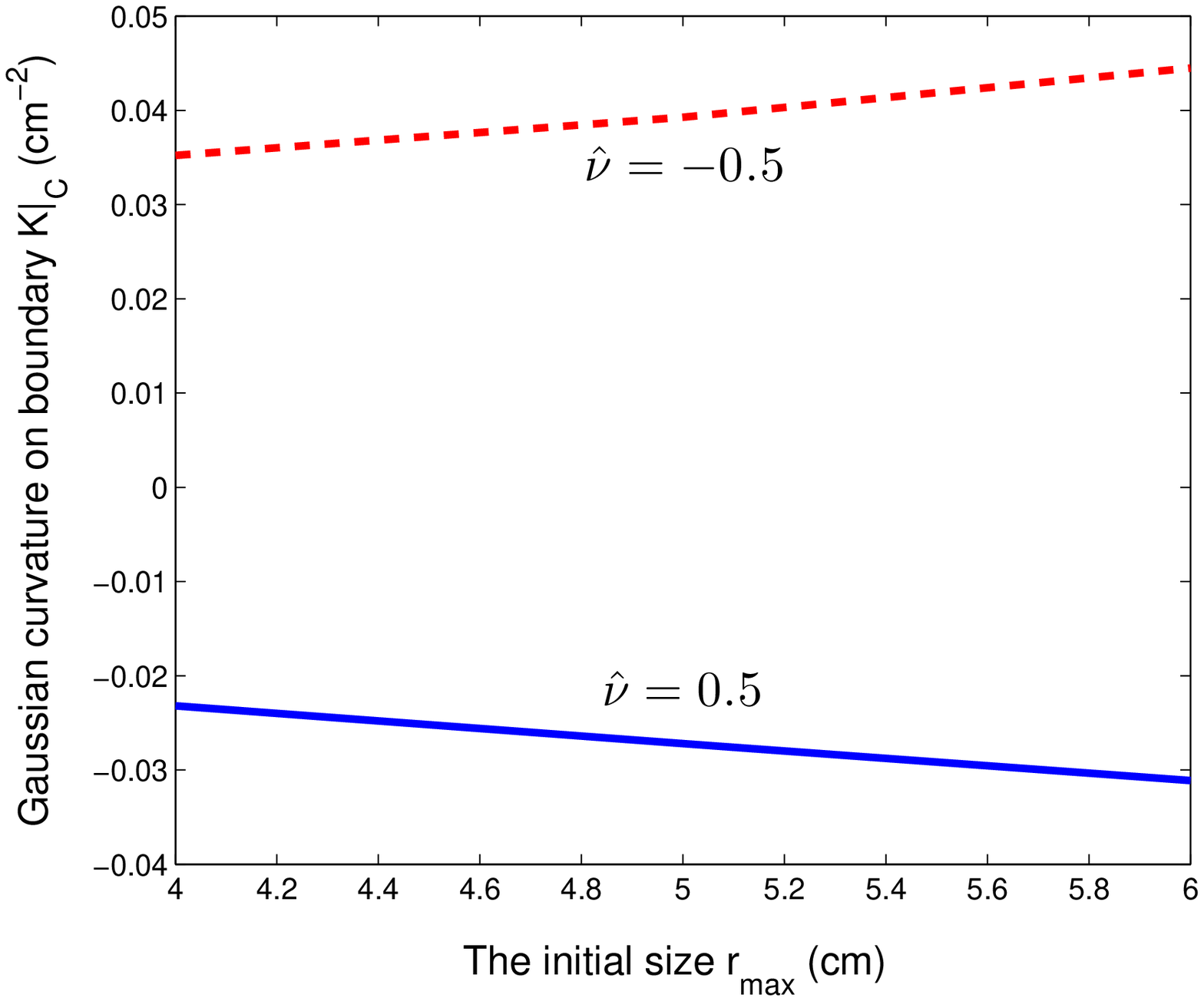}}
  \subfigure[]{
  \includegraphics[width=4.5cm]{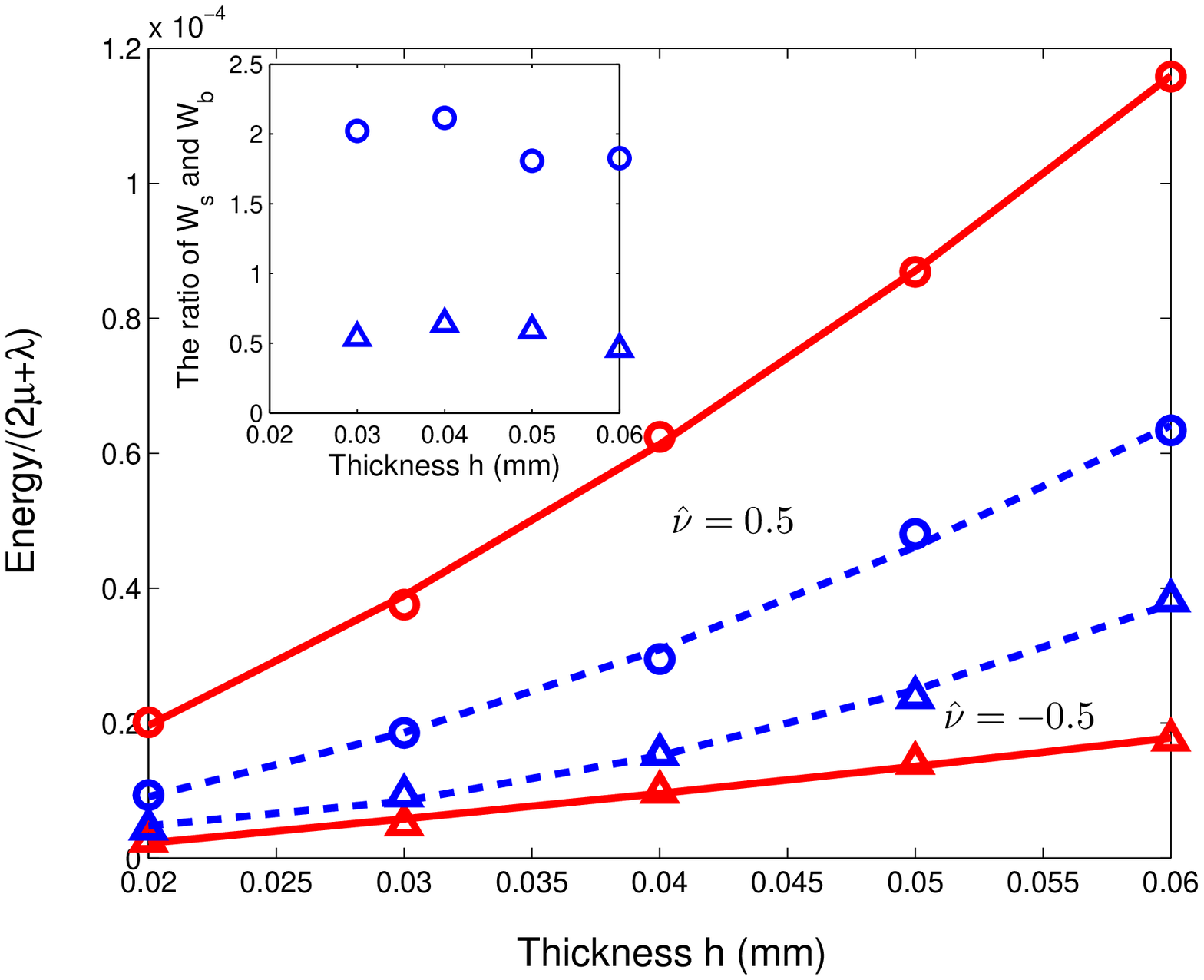}}
\caption{(a) The thickness vs. Gaussian curvature on boundary $K|_{C}$. (b) The initial size $r_{\max }$ vs. $K|_{C}$.
 (b) The in-plane stretching energy (red line) and bending energy (blue line) increase with thickness (the dot line with $\hat{\nu}=0.5$\ and the triangle line with $\hat{\nu}=-0.5$). The initial shrinkage ratio of gel sheet is $\eta _{0}=0.6-r^{2}/250$.} 
\end{figure}

\begin{figure}
\centering
\renewcommand\figurename{\small Fig.}
\subfigure[]{
  \includegraphics[width=5cm]{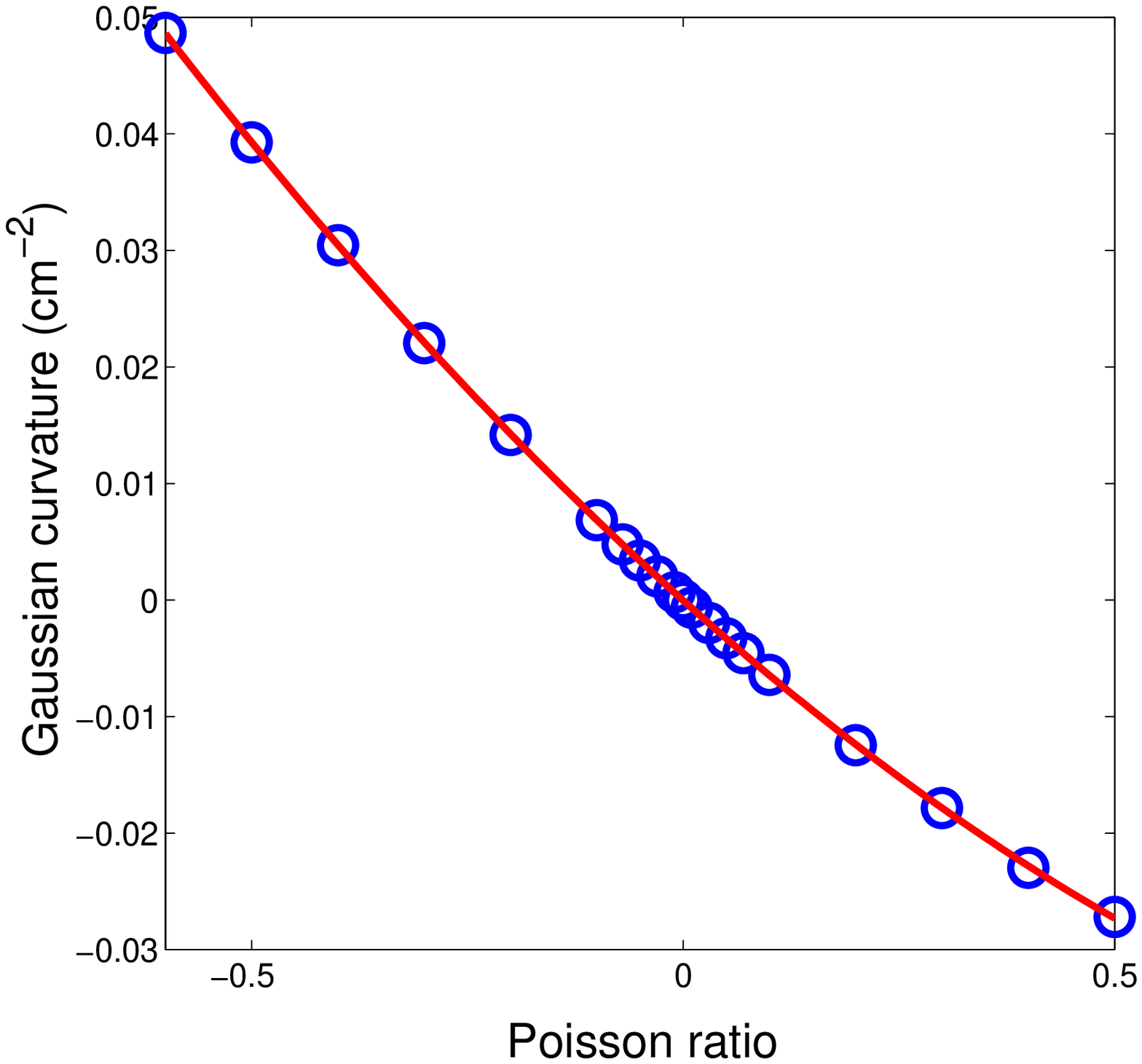}}
  \subfigure[]{
  \includegraphics[width=5cm]{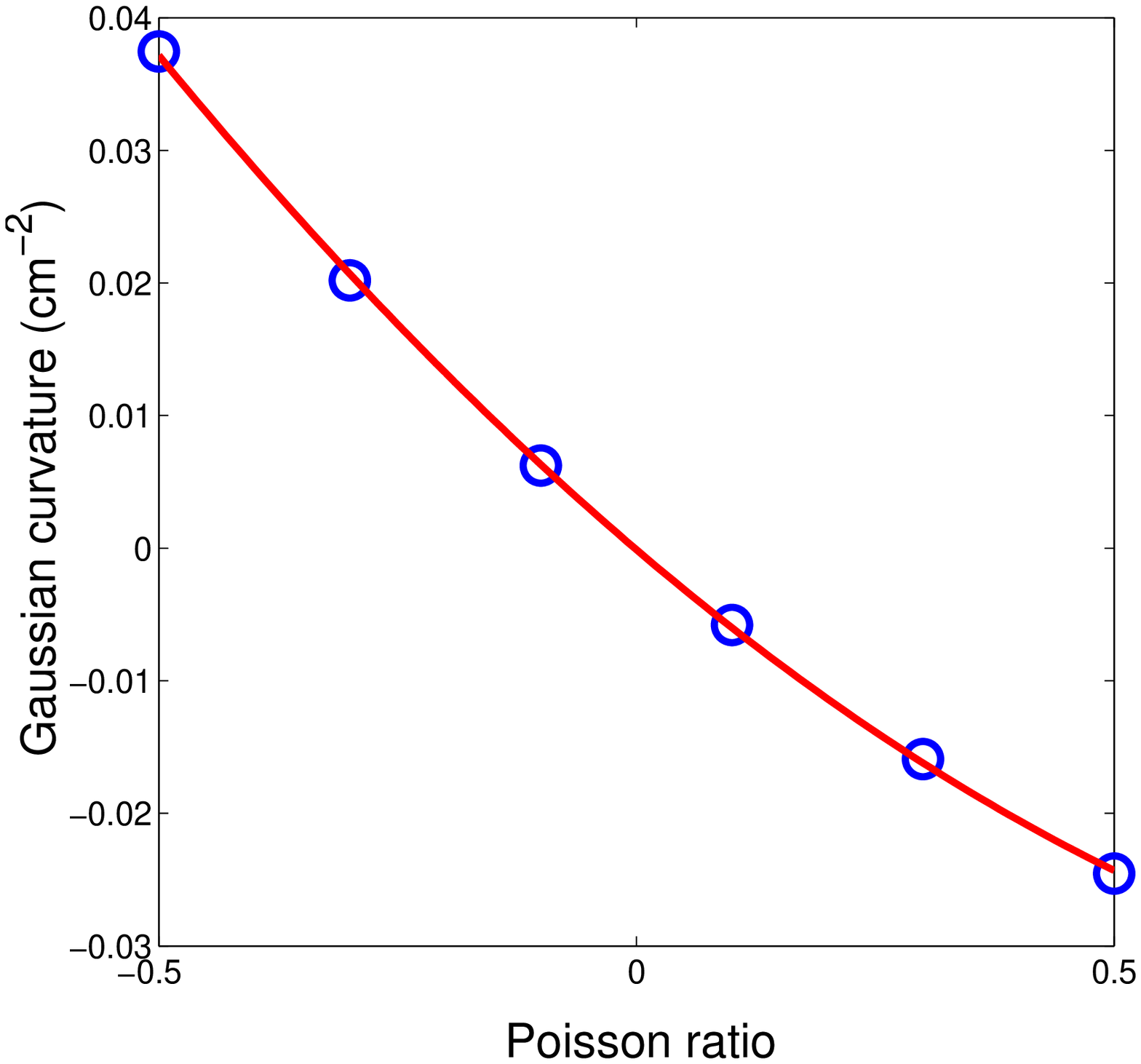}}
\caption{The Gaussian curvature on the boundary vs. Poisson ratio. (a) The dome-like sheet with $\eta _{0}=0.6-r^{2}/250$ and (b) the torus-like sheet with $\eta _{0}=0.4+1/2r^{2}$. The inset is elastic energy vs. Poisson ratio.} 
\end{figure}

Because the Poisson ratio determines the value of $c/a$, it restricts the shape near boundary. To obtain the sheets with special surface, such as sphere and cylinder, we must use special material which has the designated Poisson ratio. For the half-sphere and cylinder sheets, they need  $\hat{\nu}=0$ which derived from the boundary condition (7). However, in gel sheets the Poisson ratio is not zero generally, so the cylinder sheets made by experiment must have an additional edge in the experiment $[13]$.

\subsection{Strain}
Between the target state and equilibrium state, the sheets have different Gaussian curvature especially near boundary. So, the equilibrium state has different shape from the target state. One may expect that the sheet has a deformation near boundary. Because the target state doesn't have in-plane stretching, this deformation will induce a residual strain near boundary.

Using Eqs. (4), we obtain the strain/stress on the equilibrium shape and
obtain the density of stretching/bending energy (Fig. 4). For dome-like
sheets, the bending energy mainly distribute in the center region,
contrarily the stretching energy almost distributes near the boundary (Fig. 4a and Fig. 5a). On the center region the bending energy is greater ten
times than the stretching energy, but near boundary the bending energy is
much smaller than the stretching energy. The stress mainly locates on the
circumferential direction, and is nearly zero along radial direction. When
 $\hat{\nu}>0$, along two direction (radial and circumferential) the strains
have different sign, but when $\hat{\nu}<0$ these two strains have same
sign. For torus-like sheets, when $\hat{\nu}>0$ the $K|_{C}$ has same sign
between target and equilibrium states, so there just has a small strain near
boundary (Fig. 5b). In this case, though the in-plane stretching energy is
smaller than the bending energy, the additional stretching energy on
boundary also exist. And the additional stretching energy becomes larger
when $\hat{\nu}<0$.

\begin{figure}
\centering
\renewcommand\figurename{\small Fig.}
\subfigure[]{
  \includegraphics[width=12cm]{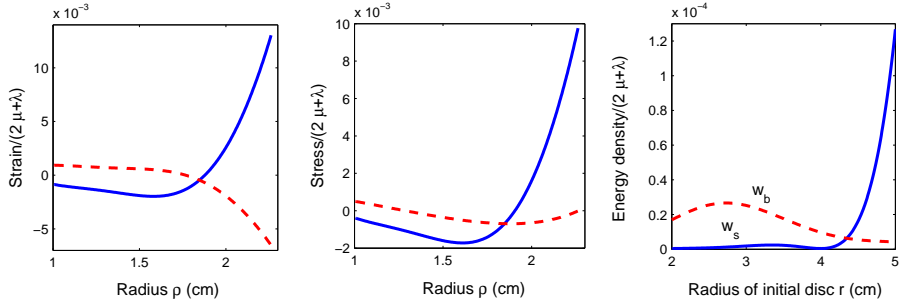}} \\
  \subfigure[]{
  \includegraphics[width=10cm]{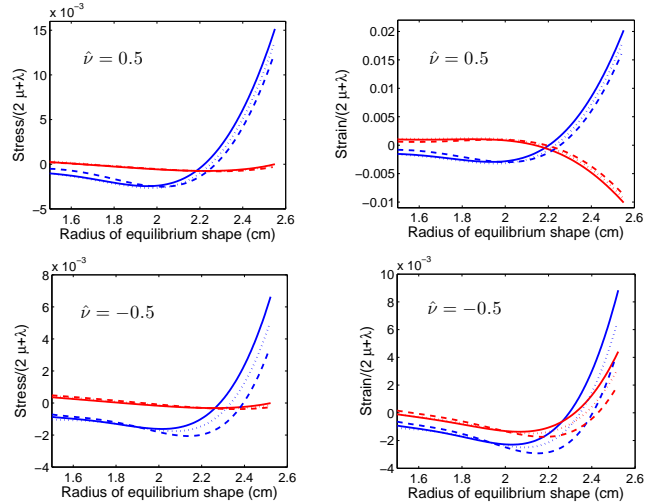}}
\caption{(a) The elastic character of the gel sheet in experiment. The first
two figures are the strain and stress distribution of equilibrium shape
along radial direction (red line) and circumferential direction (blue line).
The third figure is the distribution of bending energy (red line) and
stretching energy (blue line), respectively. In the central region, the
bending energy has a maximum value at $r=2.75$cm and the stretching energy
has a maximum value at $r=3.4$cm. Near the boundary, the stretching energy
augment rapidly. (b) The strain and stress distribution of equilibrium shape
to different boundary line tension, such as $\gamma /(2\mu +\lambda )=0$
cm (real line), $0.0005$cm (dot line), $0.001$cm (dashed). The strain and
stress along radial direction (red line) and circumferential direction (blue
line) for the gel sheet with $\eta _{0}=0.6-r^{2}/250$.} 
\end{figure}

\begin{figure}
\centering
\renewcommand\figurename{\small Fig.}
  \includegraphics[width=10cm]{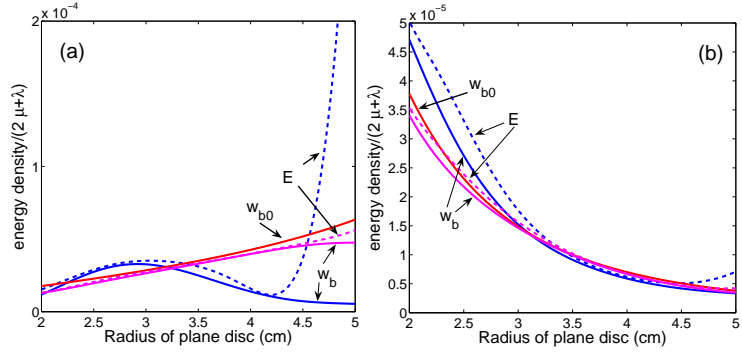}
\caption{The distribution of the elastic energy, the bending energy in
equilibrium state ($w_{b}$) and in target state ($w_{b0}$) with $\hat{\nu}=0.5$
 for (a) the dome-like sheet with $\eta _{0}=0.6-r^{2}/250$ and (b) the
minimal surface sheet with $\eta _{0}=0.4+1/2r^{2}$. The energy
distributions with boundary are blue, and without boundary are pink.} 
\end{figure}

In the experimental sheet, the border radius of equilibrium state and target
state are $2.262$cm and $2.233$cm when $\hat{\nu}=0.5$, respectively. And in
the sample D1 (we mark that the sheets with $\eta _{0}=0.6-r^{2}/250$ and
 $\eta _{0}=0.4+1/2r^{2}$ are sample D1 and sample T1, respectively), these
value are $2.571$cm and $2.500$cm, respectively. The border radius of
equilibrium state is bigger than the one of target state. To form the
negative $K|_{C}$, the boundary conditions make the sheet to bend outward
near boundary. In torus-like sheet (sample T1) with $\hat{\nu}=-0.5$, the
border radius of equilibrium state and target state is $2.081$cm and $2.100$
cm, respectively. Then, the positive $K|_{C}$ needs the sheet bending inward
near boundary. The border radius of equilibrium state is different from the
one of target state. Therefore, there is stretching on the boundary
consequentially. This stretching is reserved and gradually reduces from
boundary to center (Fig. 4). Obviously, to attain the special shape on
boundary, the sheets will acquire a deformation and have residual strain
which induces the additional stretching energy.

The ratio of bending and stretching energy is determined by Poisson ratio
(Fig. 6). In dome-like sheets, the stretching energy is larger than the
bending energy for general gels ($0.3<\hat{\nu}<0.5$). However, for the
material with $\hat{\nu}<-0.2$, in despite of the stress also cumulates near
boundary, the stretching energy change to smaller than the bending energy,
and almost disappear when $\hat{\nu}=-0.7$. There is a critical value
 $\hat{\nu}_{C}$ about the transition of domination between stretching and bending
energy. The $\hat{\nu}_{C}$ is near $-0.2$ which is close to the midpoint on
the extent of Poisson ratio $(-1\leq\hat{\nu}\leq 0.5)$. With
large Poisson ratio, the stretching energy is dominant. But, with small
Poisson ratio, the bending energy is dominant. In torus-like sheets, the
stretching energy is smaller than the bending energy. And the stretching
energy nearly disappear when $\hat{\nu}=0.5$. With large positive $\hat{\nu}$
in dome-like sheets and small negative $\hat{\nu}$ in torus-like sheets, the
stretching energy occupies a large portion of elastic energy obviously. In
these cases, the significance of boundary is very notable.

\begin{figure}
\centering
\renewcommand\figurename{\small Fig.}
\subfigure[]{
  \includegraphics[width=10cm]{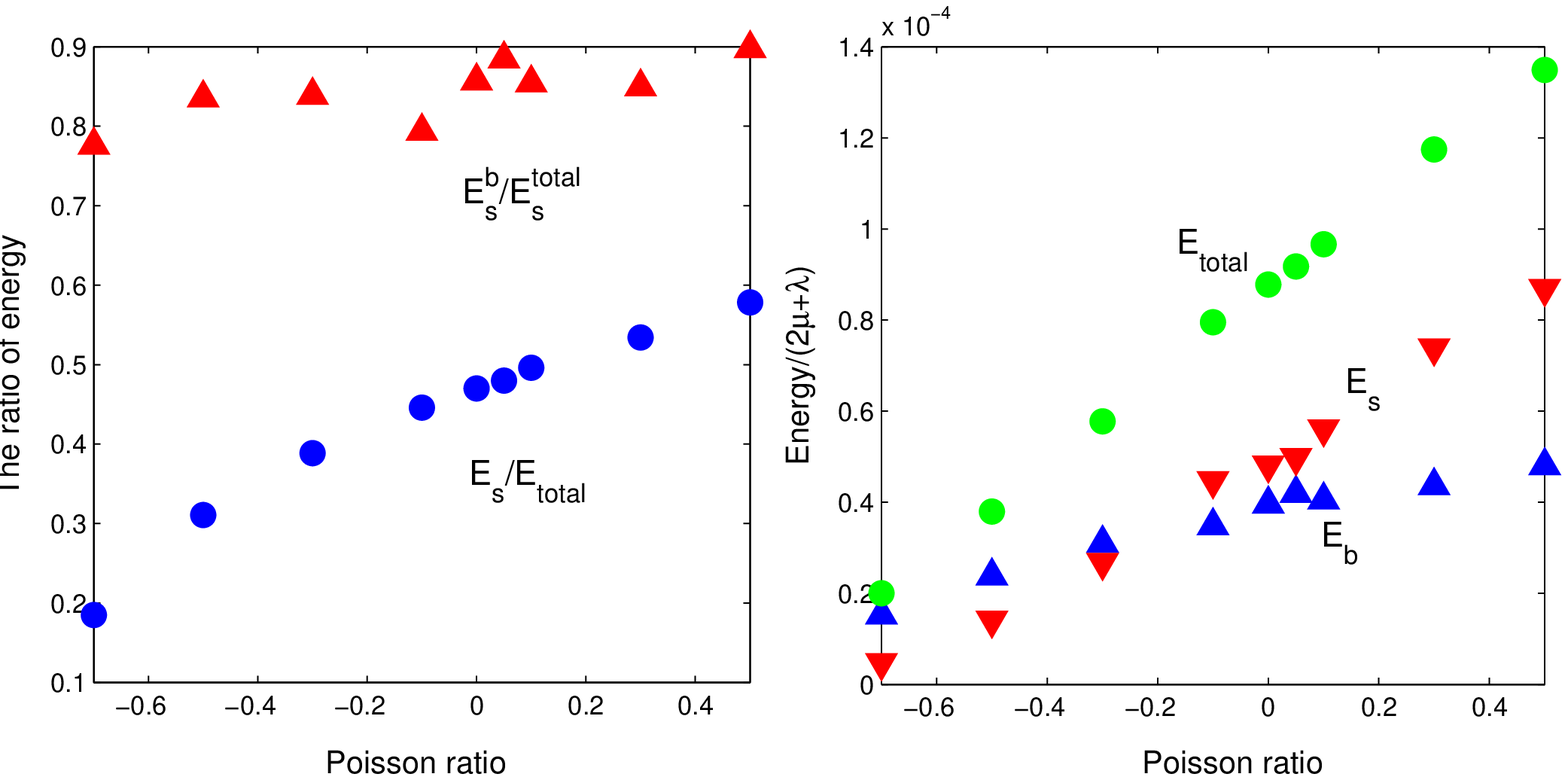}} \\
  \subfigure[]{
  \includegraphics[width=10cm]{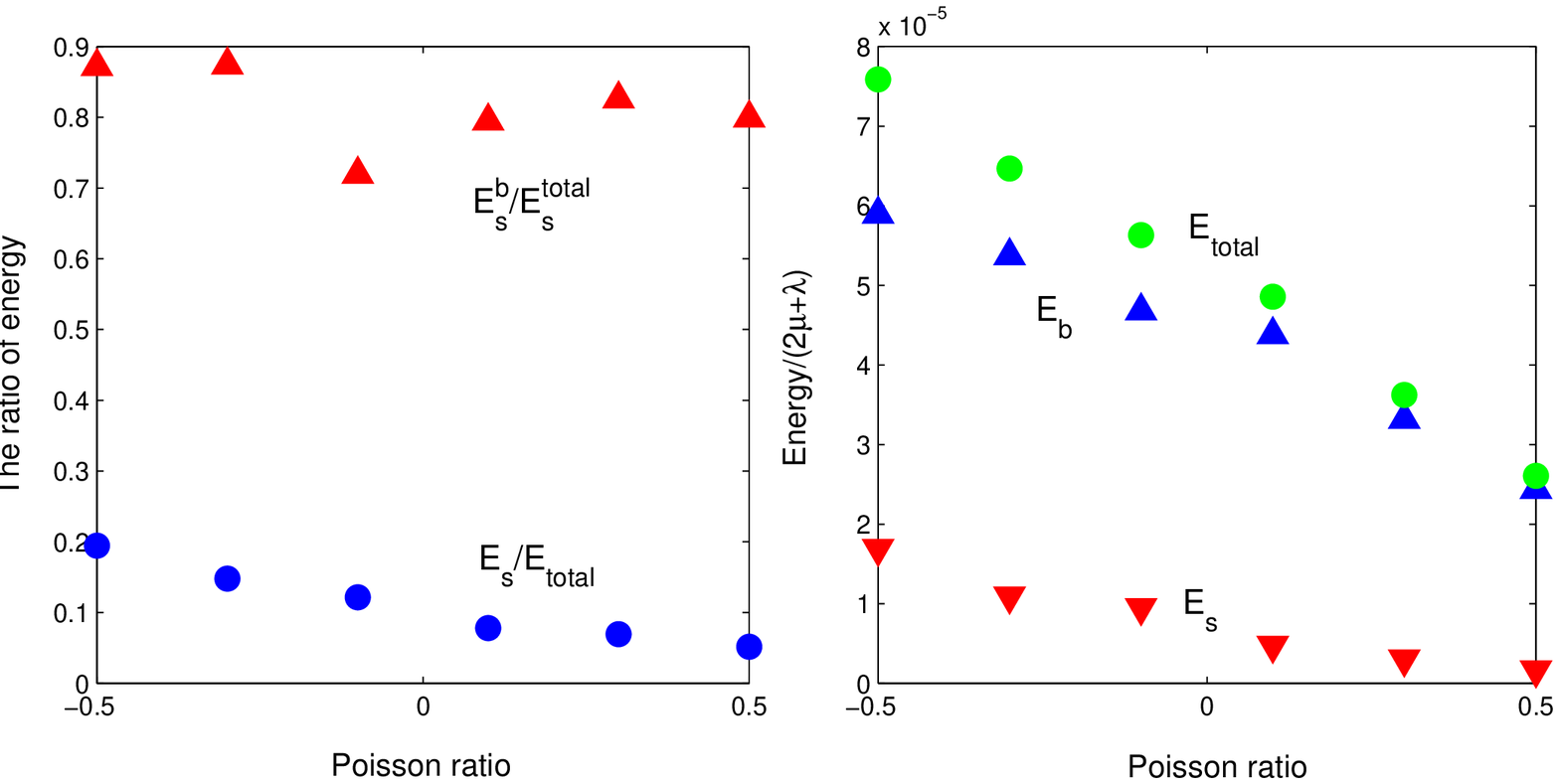}}
\caption{The stretching energy ($E_{s}$) and bending energy ($E_{b}$) vs.
Poisson ratio. The gel sheets with (a) $\eta _{0}=0.6-r^{2}/250$ and (b)
 $\eta _{0}=0.4+1/2r^{2}$. The $E_{s}^{b}$ is the stretching energy near
boundary (the outer 10\% of initial sheet). The $E_{s}^{total}$ and
 $E^{total}$ are the total stretching energy and total elastic energy.} 
\end{figure}

\subsection{The effect of boundary}
In fact, the boundary plays a decisive role to the equilibrium shape. To
discuss the role of boundary in equilibrium conformation, we compare the
sheet with and without boundary. The equilibrium shape without boundary is
different from the one with boundary generally. At cut off radius $r_{max}$
(which corresponding to the border radius of the sheet with boundary), we
label the Gaussian curvature with and without boundary as $K|_{C}^{b}$ and
 $K|_{C}^{nb}$, respectively. Comparing $K|_{C}^{b}$ and $K|_{C}^{nb}$, we
know that the boundary will change the value of Gaussian curvature. In
dome-like sheets, with $\hat{\nu}<0$ the $K|_{C}^{b}<K|_{C}^{nb}$. And with
 $\hat{\nu}>0$, the $K|_{C}^{b}$\ and $K|_{C}^{nb}$ even have opposite sign.
In torus-like sheets, with $\hat{\nu}>0$ the $K|_{C}^{b}>K|_{C}^{nb}$, and
they have opposite sign with $\hat{\nu}<0$ (Table 1). Furthermore, without
boundary the distribution of Gaussian curvature on equilibrium state just
has a little different from that on target state, however this difference is
obvious in the sheet with boundary especially near boundary (Fig. 3a).

Furthermore, a sheet with boundary has larger elastic energy than the sheet
without boundary (Table 1).\ In both dome-like and torus-like sheets, the
boundary makes the elastic energy\ augment whether $\hat{\nu}>0$ or
 $\hat{\nu}<0$. Especially, in dome-like sheet when $\hat{\nu}>0$ and torus-like sheet
when $\hat{\nu}<0$, the augment of elastic energy is notable. In dome-like
sheet this augment mainly gather near boundary and is very larger than that
in torus-like sheet which doesn't have obvious augment near boundary (Fig.
5). And, the increase of elastic energy is two times more than that in the
torus-like sheet. So, when the deformation on boundary is notable (the sign
of $K|_{C}$ between equilibrium state and target state are opposite), the
boundary will increase elastic energy obviously.

\begin{table}
\caption{The Gaussian curvature on boundary $K|_{C}$ and elastic energy with
and without boundary (the data with asterisk are with boundary, for without
boundary sheets we use the data which cut off at $r_{max}$) in
experimental sheet, dome-like sheet with $\eta _{0}=0.6-r^{2}/250$ (sample
D1), and torus-like sheet with $\eta _{0}=0.4+1/2r^{2}$ (sample T1). The
change rate $\epsilon $ is the difference of elastic energy between
equilibrium state and target state (in target state the elastic energy just
contains bending energy because the target state has no in-plane strain).}
\label{tab:1}       % Give a unique label
% For LaTeX tables use
\begin{center}
\begin{tabular}{llll} \toprule 
  & $K|_{C}$ (cm$^{-2}$) & $E/(2\mu+\lambda)$ & $\epsilon$ (\%)\\ 
experimental sheet & 0.090 & 6.83 & -7.5\\
  ($\hat{\nu}=0.5$)& -0.027* & 8.18* & 10.8* \\\hline
sample D1 & 0.120 & 10.10& -8.4 \\
 ($\hat{\nu}=0.5$) & -0.22* & 25.58* & 132.2\\\hline
sample D1 & 0.113 & 3.73& -2.5 \\
 ($\hat{\nu}=-0.5$) & 0.039* & 4.19* & 9.7 \\\hline
sample T1  & -0.036 & 3.89& -2.0 \\
 ($\hat{\nu}=0.5$)& -0.025* & 4.89* & 23.2\\\hline
sample T1 & -0.077 & 11.55& -3.1 \\
 ($\hat{\nu}=-0.5$) & 0.37* & 14.37* & 20.7\\ 
\end{tabular}
\end{center}
\end{table}

In the sheet both with and without boundary, comparing to the target state, the
stretching will reduce the bending energy. But, in the sheet with boundary,
the decrease of bending energy is more obvious. Without boundary, the
stretching makes the elastic energy of equilibrium state to be smaller than
that of the target state. However, considering the boundary, the equilibrium
state has larger elastic energy than the target state (Table 1). The
boundary is important and has a great contribution to the elastic energy.
Thus, we can find that the boundary plays an important role to format the
shape and re-distribute the stress in equilibrium conformation.

In a word, the boundary has very important effect to the strain. There is a
lot of strain accumulated near boundary. The 10\% of most outer part of
initial sheet has 90\% of total in-plane stretching energy (Fig. 6). These
boundary characters can be observed in the\ gel sheets with different
distribution of concentration and with either positive or negative Poisson
ratio. And with different initial size and thickness, the boundary always
induces the special characters of equilibrium shape and strain distribution.
The residual strain even makes the in-plane stretching energy to be larger
than the bending energy in some cases. Without boundary, the stretching will
decrease the elastic energy as close membranes. However, with boundary, the
stretching makes the elastic energy increase. Thus, the boundary has a
decisive status to determine the elastic energy. In theoretical studies
people cannot ignore the boundary.

\section{CHARACTERS OF BOUNDARY LINE TENSION}
\label{sec:3}
The theoretical discussion of equilibrium shape didn't consider the boundary
line tension in the previous studies $[17,18,21]$. However, the boundary
term $E_{C}$ is an important component of conformation energy $[24,27]$. In
this section, we will consider the boundary line tension and discuss its
effect.

\subsection{Border radius}
Using the equilibrium shape equations (Eqs. (6)), we obtain the equilibrium
shape with different boundary line tension. In dome-like sheets, the
boundary line tension and the border radius has a linear relationship for
both positive and negative Poisson ratio (Fig. 7). The similar linear
relationship also can be found in the torus-like sheets. The border radius
of equilibrium shape will decrease when the boundary line tension augment.
So the boundary line tension will promotes the shrinkage on the edge. The
border radius is easily to be measured in the experiment. Its theoretic
value can be compared with the experimental data in straightway. In
experiment, the accurate boundary line tension is measured hardly.
Fortunately, the linear relationship between boundary line tension and
border radius can offers a method to measure the boundary line tension by
the border radius which can be obtained directly.

\begin{figure}
\centering
\subfigure[]{
  \includegraphics[width=5cm]{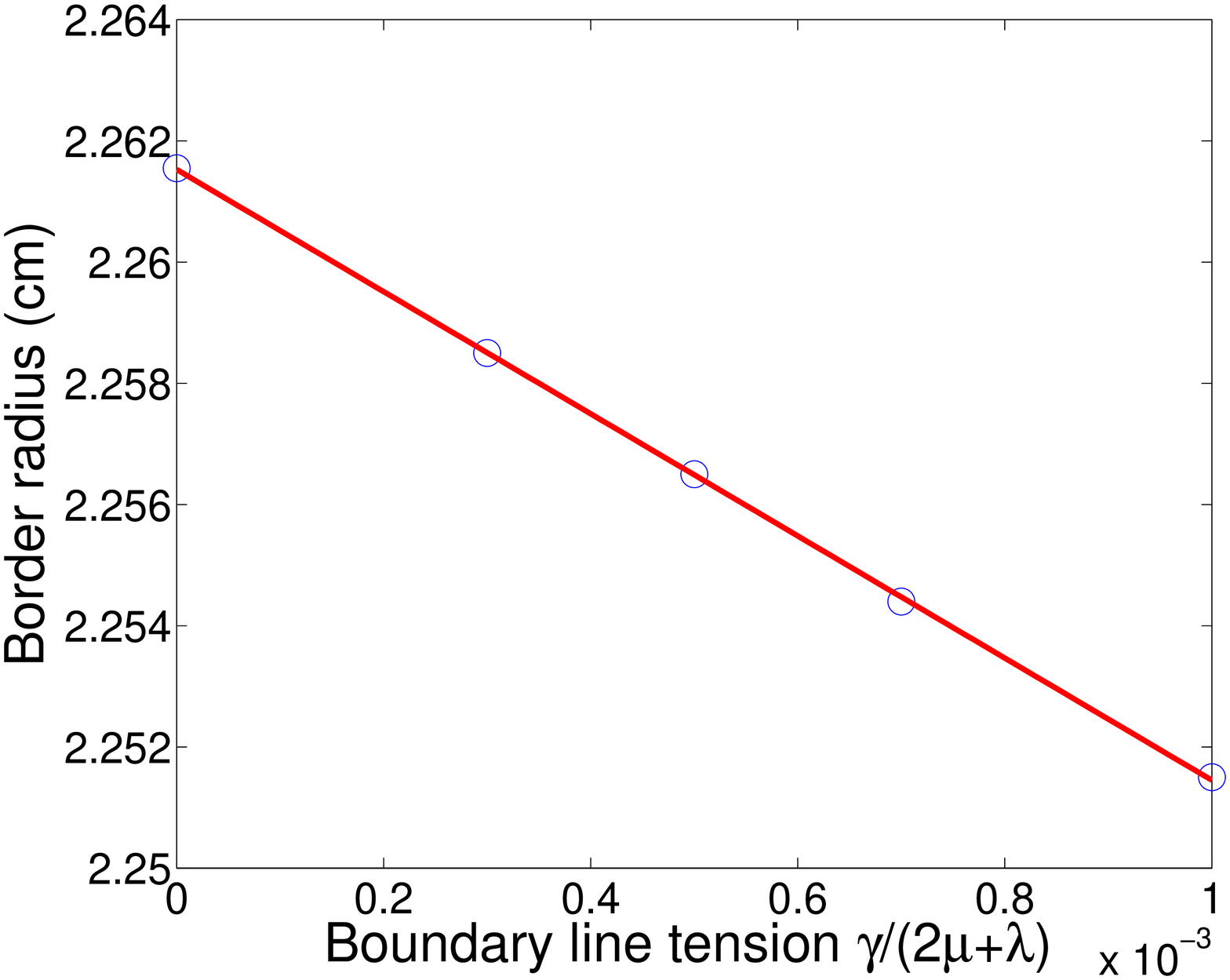}}
  \subfigure[]{
  \includegraphics[width=5cm]{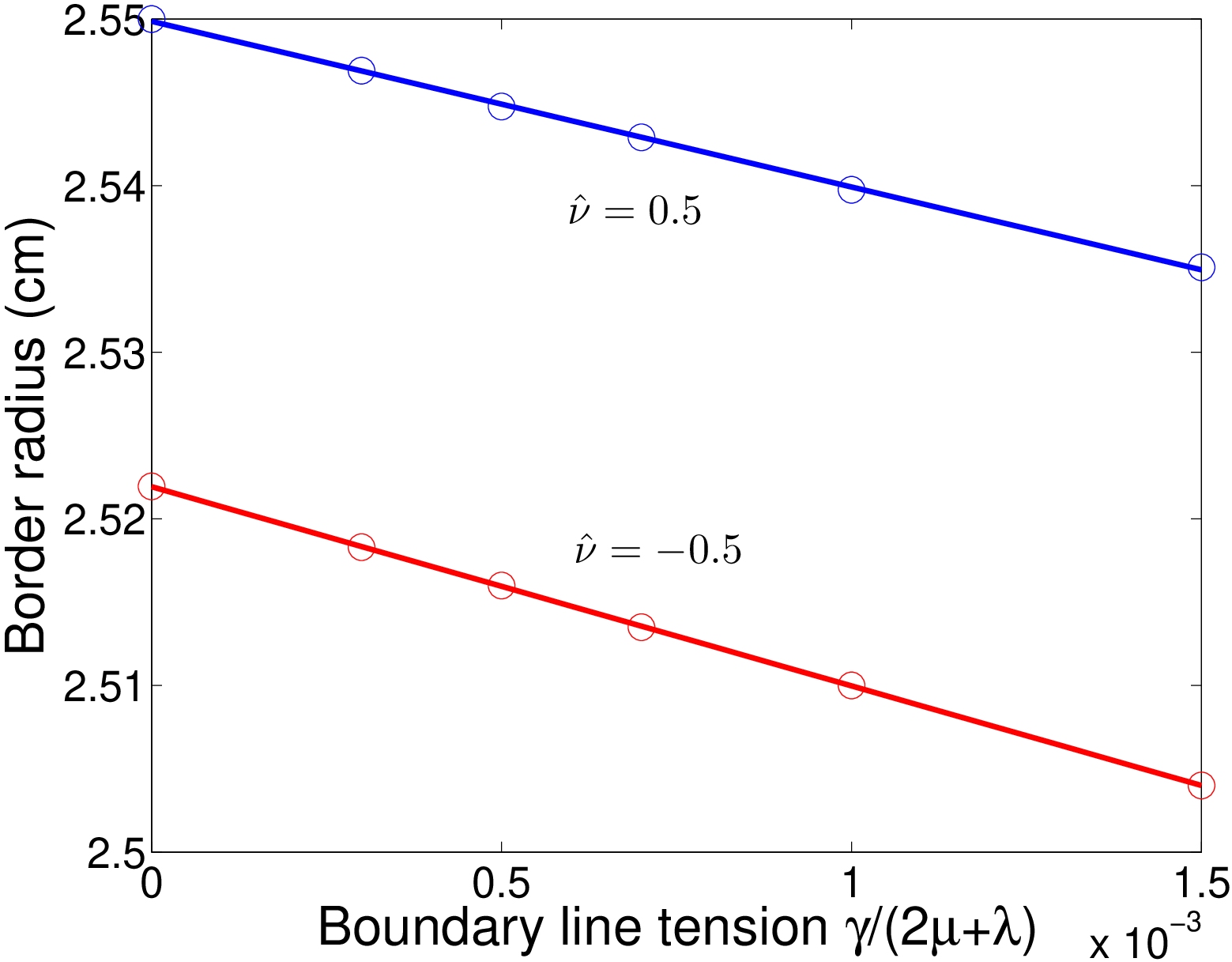}}
\caption{(a) Boundary line tension vs. border radius for the experimental
sheet. (b) The relation between boundary line tension and border radius for
$\hat{\nu}=0.5$ (blue line) and $\hat{\nu}=-0.5$ (red line) in the gel sheets
with $\eta _{0}=0.6-r^{2}/250$.} \label{fig:7}
\end{figure}

From the experiment observation $[13]$, the border radius of gel sheet is
 $2.2584$cm. However, it is $2.262$cm from the theoretical calculation
with no boundary line tension $[18]$. Apparently, the theoretical border
radius with no boundary line tension is bigger than the experimental
observation data. For precise calculation, the boundary line tension is
needed. The theoretical border radius is same as the experimental data
 ($2.2584$cm) in case of $\gamma /(2\mu +\lambda )=0.00031$cm (Fig.
7a). Considering the Young modulus of NIPA gel is $0.11$MPa $[30]$ and
from Eqs. (2), it implies that the corresponding boundary line tension is
 $2.3\times 10^{-5}$J/cm. Furthermore, the residual strains on boundary with
and without $\gamma $ are $0.0132$ and $0.0116$, respectively. We know
that the residual strain decides the equilibrium conformation of gel sheet.
The elastic energy $F/(2\mu +\lambda )=7.547\times 10^{-5}$ and $8.180\times
10^{-5}$ for with and without $\gamma $, respectively. So, although the
difference of border radius between with and without $\gamma $ is small,
the $\gamma $ also has a big influence to the equilibrium conformation.

\subsection{Strain}
The different boundary line tension just affects the magnitude of Gaussian
curvature and does not change the sign of $K$ on boundary (Fig. 1a). In the
dome-like sheets, when $\hat{\nu}>0$ the strain and stress almost distribute
near the boundary (Fig. 4). For small boundary line tension, it will weaken
the in-plane stretching but just have less influence to the distribution of
strain and stress (Fig. 4b).

The bending energy weakly enhances with boundary line tension $\gamma $
(Fig. 8). For rotational symmetric sheets, the boundary term of conformation
energy is $E_{C}=\oint_{C}\gamma ds=2\pi \gamma R_{C}$. With large boundary
line tension $\gamma $, the border radius $R_{C}$ will tend smaller to
reduce the energy of $E_{C}$. Along the boundary, the displacement is
 $R_{C}-\tilde{R}_{C}$, so the stretching energy is proportion to
  $|R_{C}-\tilde{R}_{C}|$. And because the $\gamma $ will decrease $R_{C}$, the large $\gamma $
weakens the stretching energy when $R_{C}>\tilde{R}_{C}$. But when $R_{C}$
becomes smaller than $\tilde{R}_{C}$ the $\gamma $ will make the stretching
energy to increase. In dome-like sheets, when $\hat{\nu}>0$ the boundary
need to spread outward. The bending energy is increase with $\gamma $
increasing. But the in-plane stretching energy is decrease with\ $\gamma $
increasing. There is a critical point at $\gamma /(2\mu +\lambda )=0.001$ in
which point the domination between stretching and bending energy is
transition. When $\hat{\nu}<0$, with\ $\gamma $ increasing the in-plane
stretching energy decrease firstly and increase when $\gamma /(2\mu +\lambda
)>0.001$ (Fig. 8). In the torus-like sheet, with $\gamma =0$\ the $R_{C}$ is
smaller than $\tilde{R}_{C}$ for both $\hat{\nu}>0$ and $\hat{\nu}<0$. Thus,
the in-plane stretching energy will increase with $\gamma $ increasing.

\begin{figure}
\centering
\renewcommand\figurename{\small Fig.}
\subfigure[]{
  \includegraphics[width=5cm]{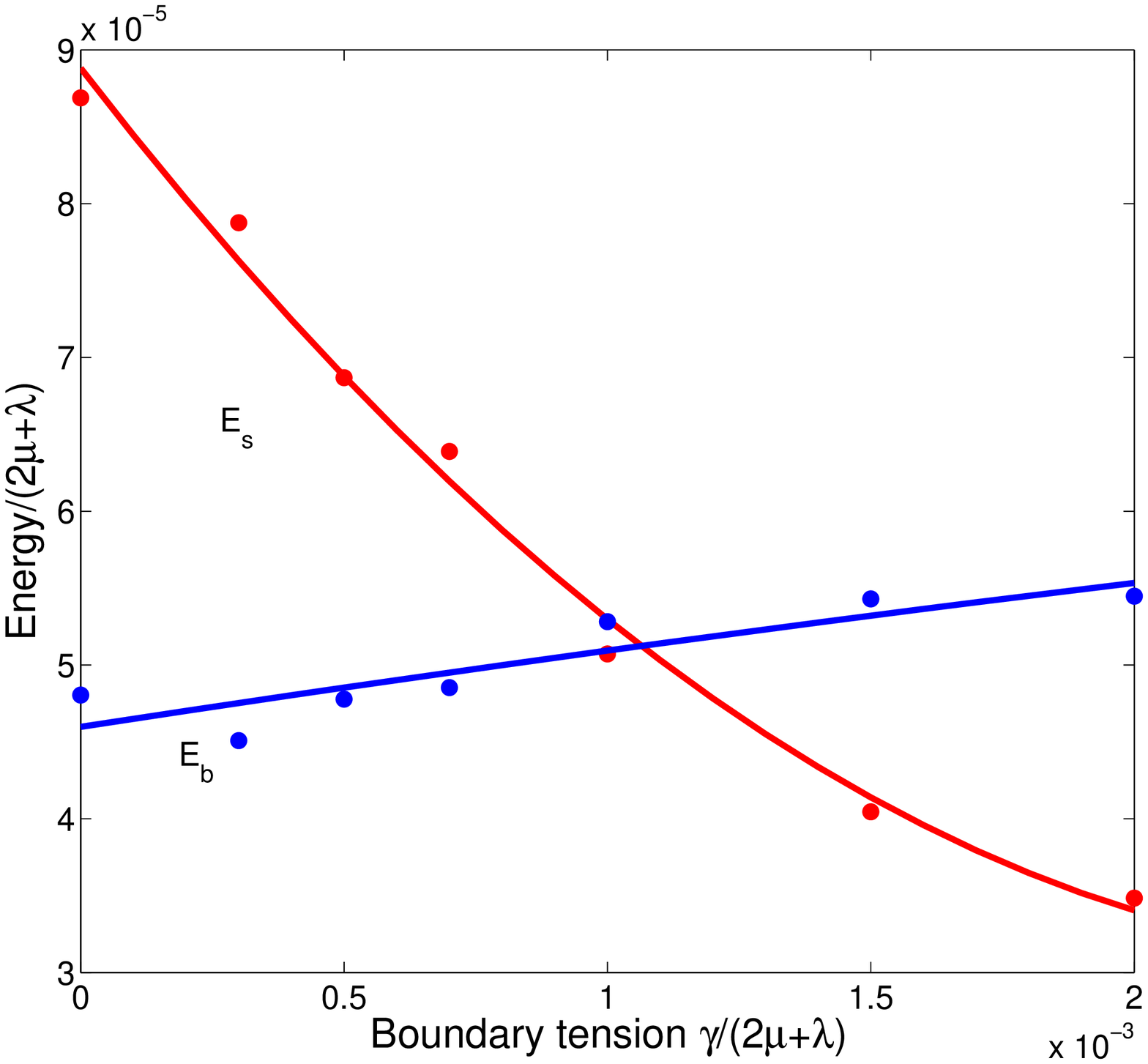}}
  \subfigure[]{
  \includegraphics[width=5cm]{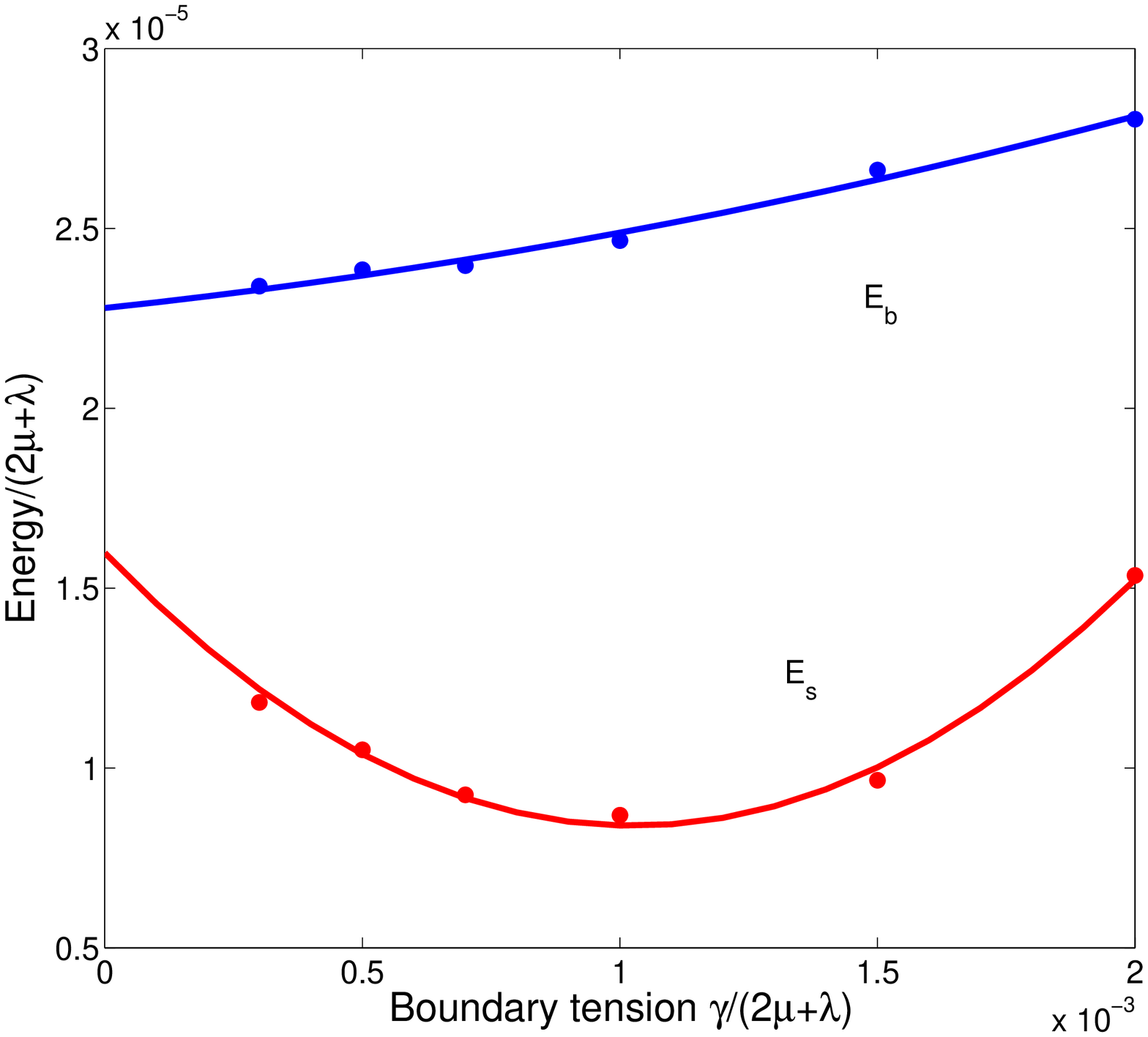}}
\caption{The boundary line tension vs. the stretching energy $E_{s}$ (red line)
and bending energy $E_{b}$ (blue line). The gel sheets with $\eta
_{0}=0.6-r^{2}/250$ when (a) $\hat{\nu}=0.5$ and (b) $\hat{\nu}=-0.5$.} \label{fig:8}
\end{figure}

In a word, the boundary line tension has obvious effect on the elastic
energy. So, in exact studies of gel sheets, $\gamma $ need be considered.
But, $\gamma $ doesn't change the type of equilibrium shape and just affects
the value of elastic energy and the size of equilibrium shape. Thus, the
special mechanical and geometrical characteristic on boundary is not caused
by boundary line tension. In other word, the boundary characters are indeed
induced by the residual strain on boundary.

\section{CONCLUSION}
\label{sec:4}
In this paper, we investigate the role of boundary on equilibrium
conformation and study the effect of boundary line tension $\gamma $. We
find that the boundary can not be ignored in gel sheets. The boundary
affects the competition of bending and stretching in the equilibrium
conformation of gel sheets. For the rotational symmetric gel sheets, the
boundary has special Gaussian curvature which is opposite to $\hat{\nu}$.
This induces a deformation and residual strain near boundary. Consequently,
the 10\% of most outer part of initial sheet has almost 90\% of total
in-plane stretching energy. In some cases, the residual strain even makes
the in-plane stretching energy larger than the bending energy.

The boundary plays an essential role in the equilibrium conformation. With
or without boundary, the equilibrium shapes are different obviously. The
distribution of Gaussian curvature is different in both the situations.
Furthermore, without boundary, the stretching will decrease the elastic
energy as closed membranes. However, the boundary will increase the elastic
energy. With boundary the elastic energy of equilibrium state is larger than
that of target state in some cases. These phenomena are resulted from the
residual strain on boundary.

To agree well with experimental result, in theoretical studies we find that
the boundary line tension $\gamma $ must be involved. $\gamma $ impacts the
residual strain on boundary. Then, considering $\gamma $ or not, the
elastic energies have difference about 10\%. Generally, the $\gamma $
decreases the elastic energy. The boundary line tension and border radius
are one-to-one correspondence, therefore there is a simple way to measure
the boundary line tension by the border radius.

\section*{acknowledgement}
We thank M. G. Xia, E. H. Zhang, and D. S. Lei for helpful
discussions. This work was supported by the Cultivation Fund of the
Key Scientific and Technical Innovation Project, Ministry of
Education of China (NO. 708082) and Chinese NSF grant NO. 11074196.

\end{document}